\documentclass[twocolumn,floats,showpacs,amsmath,amssymb,prb]{revtex4}
\usepackage{graphicx}

\begin{document}

\title{Molar volume of solid isotopic helium mixtures}
\author{Carlos P. Herrero}
\affiliation{Instituto de Ciencia de Materiales,
         Consejo Superior de Investigaciones Cient\'{\i}ficas (CSIC), 
         Campus de Cantoblanco, 28049 Madrid, Spain }
\date{\today}

\begin{abstract}
 Solid isotopic helium mixtures have been studied by path-integral 
Monte Carlo simulations in the isothermal-isobaric ensemble.
This method allowed us to study the molar volume as a 
function of temperature, pressure, and isotopic composition.
At 25 K and 0.2 GPa, the relative difference between molar volumes 
of isotopically-pure crystals of $^3$He and $^4$He is found to be 
about 3\%. This difference decreases under pressure, and for 12 GPa it 
is smaller than 1\%. For isotopically-mixed crystals, 
a linear relation between lattice parameters and concentrations
of helium isotopes is found, in agreement with Vegard's law. 
The virtual crystal approximation, valid for isotopic mixtures of 
heavier atoms, does not give reliable results for solid solutions of 
helium isotopes.
\end{abstract}



\maketitle

\section{Introduction}

 The lattice parameters of two chemically identical crystals with different
isotopic composition are not equal, lighter isotopes giving rise to 
larger lattice parameters. This is due to the dependence of atomic 
vibrational amplitudes upon atomic mass, along with the anharmonicity of 
the vibrations.
This effect is most important at low temperatures, since the zero-point
amplitude decreases with increasing atomic mass.
At higher temperatures, the isotope effect on the crystal
volume is less relevant, and disappears in the high-temperature
(classical) limit at $T > \Theta_D$ ($\Theta_D$, Debye temperature),
where vibrational amplitudes become independent of the atomic mass.
At present, the isotopic effect in the lattice parameters
of crystals can be measured with high precision \cite{ka98}.

Other quantities, such as the vibrational energy, display an isotope 
dependence at low $T$ in a harmonic approximation, due
to the usual rescaling of the phonon frequencies with the
isotopic mass $M$ ($\omega \propto M^{-1/2}$),
but this dependence can show appreciable changes when
anharmonic effects are present.
All these effects are expected to be more important in the
case of helium than for heavier atoms.
In fact, solid helium is in many respects an archetypal
``quantum solid,'' where zero-point energy and associated anharmonic 
effects are appreciably larger than in most known solids.
This gives rise to peculiar properties, whose understanding has presented 
a challence for theories and modelling from a microscopic point of view
\cite{ce95}. 

Anharmonic effects in solids, and in solid helium in particular,
have been studied theoretically for many years by using techniques such
as quasiharmonic approximations and self-consistent phonon
theories \cite{kl76,sr90,va98}.
 An alternative procedure is based on the Feynman path-integral formulation
of statistical mechanics \cite{fe72,kl90}, that has turned out to be a
convenient approach to study thermodynamic 
properties of solids at temperatures lower than their Debye temperature
$\Theta_D$, where the quantum character of the atomic nuclei is relevant.
In particular, Monte Carlo or molecular dynamics sampling can be 
applied to evaluate finite-temperature path integrals, thus allowing one to 
carry out quantitative and non-perturbative studies of anharmonic effects 
in solids \cite{ce95}.

 The path-integral Monte Carlo (PIMC) method has been employed to study
several structural and thermodynamic properties of solid 
helium \cite{ce95,ce96,ba89,bo94,ch01}, as well
as heavier rare-gas solids \cite{cu93,mu95,ch02,ne02,he02,he03a}.
For helium, in particular, this procedure has predicted kinetic-energy
values \cite{ce96} and Debye-Waller factors \cite{dr00} in good agreement
with data derived from experiments \cite{ar03,ve03}.
PIMC simulations have been also performed to study the isotopic shift
in the helium melting pressure \cite{ba89,bo94}.

In most calculations of properties of crystals with isotopically mixed
composition, it is usually assumed that each atomic nuclei in the solid
has a mass equal to the average mass. This kind of {\em virtual-crystal
approximation} has been used in density-functional calculations, as well
as in PIMC simulations \cite{de96,ca00,zh00,he00a,he01b,he03a}. 
In fact, in earlier simulations it was found
that the results obtained by using this approximation are indistinguishable
from those derived from simulations in which actual isotopic mixtures 
were considered. This seems to be true for atoms heavier than helium, and
in particular for rare-gas solids including solid Ne \cite{he03a}, but is not 
guaranteed to happen for solid helium, due to its low atomic mass and
large anharmonicity.

It is well kown that at temperatures lower than 1 K, a phase separation
appears in solid $^3$He-$^4$He mixtures, and the actual temperature at which
this separation occurs depends on pressure and isotopic composition 
\cite{ar80,su82}. This isotope segregation is due to the different molar 
volume of both isotopes, which in turn is caused by the different zero-point 
vibrational amplitudes of $^3$He and $^4$He \cite{mu68,su82}.
In this paper, we consider mixtures of helium isotopes at higher
temperatures, where $^3$He and $^4$He form solid solutions for any isotopic
composition. By varying the molar fraction of both isotopes, we
analyze changes in the lattice parameter and kinetic energy, by using 
PIMC simulations.  We employ the isothermal-isobaric ($NPT$) ensemble,
which allows us to consider properties of these solid solutions along 
well-defined isobars.
This simulation method permits to study properties of actual isotopic 
mixtures, and compare them with those obtained for virtual crystals in which
each atom has a mass equal to the average mass of the considered 
isotope mixture.

The paper is organized as follows.  In Sec.\,II, the
computational method is described. In Sec.\,III we present the results,
and  Sec.\,IV includes a discussion and the conclusions.

\section{Method}
                                                                                    
Equilibrium properties of solid helium in the face-centred cubic (fcc) and
hexagonal close-packed (hcp) phases have been calculated by PIMC simulations 
in the $NPT$ ensemble.
Our simulations were performed on supercells of the fcc and hcp
unit cells, including 500 and 432 atoms respectively.
These supercell sizes are enough for convergence of the quantities
studied here \cite{he06}.
For a given average isotopic mass $\overline{M}$, we randomly distribute 
$^3$He and $^4$He atoms in the appropriate proportions over the lattice sites of 
the simulation cell, and the atoms are kept fixed at their respective positions 
along the simulation (no diffusive positional changes).
This is assumed to be valid for the temperatures studied here, much higher
than those at which phase separation appears ($T <$ 1 K) \cite{ar80,su82}.
For each isotopic composition studied here, we have taken five different 
realizations of the isotope mixture. To analyze with more detail the
dispersion in the lattice parameters obtained in the simulations, we took
12 different samples for $\overline{M} = 3.25$ amu, and 15 samples for 
$\overline{M} = 3.5$ amu (see below).
For comparison, we have also considered virtual crystals, in which every atom
has a mass $\overline{M}$, i.e., $M_i = \overline{M}$ for $i = 1, ..., N$.

Helium atoms have been considered as quantum particles interacting
through an effective interatomic potential, composed of a two-body
and a three-body part.  For the two-body interaction, we have
taken the potential developed by Aziz {\em et al.} \cite{az95}
(the so-called HFD-B3-FCI1 potential). For the three-body
part we have employed a Bruch-McGee-type potential \cite{br73,lo87},
with the parameters given by Loubeyre \cite{lo87}, but with
parameter $A$ in the attractive exchange interaction rescaled by a 
factor 2/3, as in Ref. \cite{bo94}.
This interatomic potential has been found to describe well the vibrational 
energy and equation-of-state of solid helium in the available range of 
experimental data, including pressures on the order of 50 GPa \cite{he06}.

The PIMC method relies on an isomorphism between the considered quantum 
system and a classical one, obtained by replacing each quantum particle 
by a cyclic chain of $N_{Tr}$ classical particles 
($N_{Tr}$: Trotter number), connected
by harmonic springs with a temperature-dependent constant.
This isomorphism appears because of a discretization of the density
matrix along cyclic paths, that is usual in the path-integral formulation
of statistical mechanics \cite{fe72,kl90}.
Details on this computational method can be found
elsewhere \cite{gi88,ce95,no96}.

Our simulations were based on the so-called ``primitive'' form
of PIMC \cite{ch81,si88}.
We considered explicitly two- and three-body terms in the simulations,
which did not allow us to employ effective forms for the density matrix,
developed to appreciably simplify the calculation when only two-body
terms are explicitly considered \cite{bo94}.
Quantum exchange effects between atomic nuclei were not taken into account,
since they are negligible for solid helium at the temperatures and pressures
studied here. (This is expected
to be valid as long as there are no vacancies and $T$ is higher than the
exchange frequency $\sim 10^{-6}$ K \cite{ce95}.)
For the energy we have used the ``crude'' estimator, as defined in 
Refs. \cite{ch81,si88}.

Sampling of the configuration space has been carried out by the Metropolis
method at pressures $P \leq$ 12 GPa, and temperatures between 10 K and the
melting temperature at each considered pressure.
However, most of the simulations presented in this paper were carried out 
for fcc He at 25 K and 0.3 GPa,
conditions at which the isotopic effects studied here are clearly 
observable. Some simulations were also carried out at lower temperatures
(see below).
 For given temperature and pressure, a typical run consisted
of $10^4$ Monte Carlo steps for system equilibration, followed by $10^5$ 
steps for the calculation of ensemble average properties.
To keep roughly constant the accuracy of the computed quantities
at different temperatures, we have taken a Trotter number that
scales as the inverse temperature $1/T$. At a given $T$, the 
value of $N_{Tr}$ required to reach convergence of the results depends
on the Debye temperature, higher $\Theta_D$ requiring larger 
$N_{Tr}$.  Since vibrational frequencies 
increase as the applied pressure rises, $N_{Tr}$ has to be raised
accordingly.  For pressures on the order of 1 GPa, $N_{Tr} T$ = 2000 K is
enough to reach convergence of the computed quantities. For pressures larger
than 2 GPa, we have taken $N_{Tr} T$ = 4000 K for $^3$He and 3000 K for 
$^4$He, as in earlier work \cite{he06}.
Other technical details are the same as those employed in Refs.
\cite{no97,he02,he06}.

In the isothermal-isobaric ensemble, the mean-square fluctuations
in the volume $V$ of the simulation cell are given by
\begin{equation}
      \sigma_V^2 = \frac{V}{B} k_B T   \; ,
\label{dv2}
\end{equation}
where $B = - V ( {\partial P} / {\partial V} )_T$
is the isothermal bulk modulus \cite{la80}. Hence, for a cubic crystal, 
the fluctuations in the lattice parameter $a$ are:
\begin{equation}
      \sigma_a^2 = \frac{k_B T}{9 L^3 a B}   \; ,
\label{da2}
\end{equation}
where $L^3$ is the number of unit cells in a simulation cell with
side length $L a$.
From Eq.\,(\ref{dv2}) one can see that the relative fluctuation
in the volume of the simulation cell, $\sigma_V / V$, scales as 
$L^{-3/2}$.  For $^4$He at 25 K we found in PIMC simulations 
$\sigma_V / V = 5.5 \times 10^{-3}$ for $P$ = 0.3 GPa, 
and $1.7 \times 10^{-3}$ for 12 GPa (for $L=5$ and 500 atoms). 
For fcc $^4$He at 0.3 GPa this
translates into $\sigma_a = 7.3 \times 10^{-3}$ \AA.
We are interested in an accuracy in $a$ on the order of $10^{-4}$ \AA,
which means that at this temperature and pressure one needs about
$10^4$ independent data.  In practice, the required number
of Monte Carlo steps is expected to be larger, due to correlation between
configurations along a Monte Carlo trajectory.
In fact, we have checked that $10^5$ Monte Carlo steps are enough to have 
an statistical uncertainty in $a$ on the order of $10^{-4}$ \AA.

\section{Results}

We have checked the convergence with the Trotter number of several
quantities derived from our PIMC simulations. 
In Fig.~1 we display the dependence of the lattice parameter $a$ of
fcc $^3$He and $^4$He, as a function of the inverse Trotter number
$N_{Tr}^{-1}$, for $T$ = 25 K and $P$ = 0.3 GPa. The lattice parameter
obtained in the simulations increases with $N_{Tr}$, and converges to
a finite value for large $N_{Tr}$ (limit $N_{Tr}^{-1} \to 0$ in Fig.~1).
The difference $\delta a = a_3 - a_4$ between lattice parameters of
$^3$He and $^4$He decreases as the Trotter number is lowered, and
goes to zero in the classical limit ($N_{Tr} = 1$), where this isotopic
effect disappears.
The reliability of the interatomic potential employed here to predict
lattice parameters of solid helium has been studied elsewhere \cite{he06}. 
Here we will only comment that it gives good agreement with experimental
results up to the melting temperature in the pressure range considered
in this paper. Thus, for fcc $^4$He at 38.5 K and 0.493 GPa we find
$a$ = 3.9154 \AA\ vs 3.915(2) \AA\ derived from inelastic neutron 
scattering \cite{th78}.

\begin{figure}
\vspace{-2.0cm}
\includegraphics[width= 9cm]{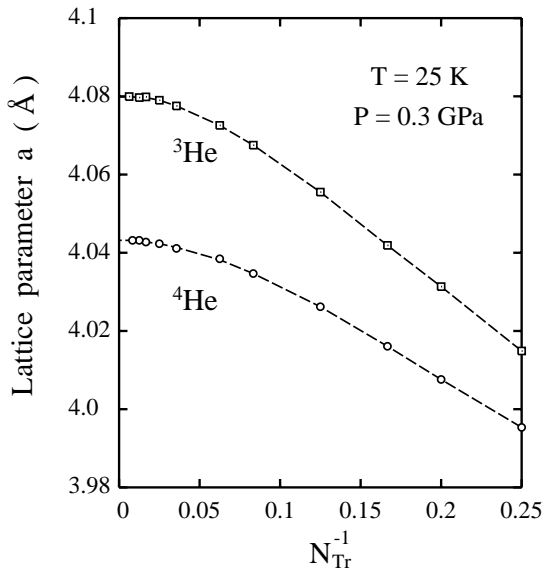}
\vspace{-2.5cm}
\caption{\label{f1}
Convergence of the lattice parameter $a$ of fcc He as a function
of the inverse Trotter number, $N_{Tr}^{-1}$, as derived from PIMC
simulations at $T$ = 25 K and $P$ = 0.3 GPa. Squares and circles
correspond to $^3$He and $^4$He, respectively. Dashed lines are guides to
the eye. Error bars are smaller than the symbol size.
}
\end{figure}

Once checked its convergence with $N_{Tr}$,
in Fig.~2 we show the temperature dependence of the lattice parameter
$a$ of fcc $^3$He and $^4$He at two different pressures: 0.3 GPa (open
symbols) and 0.6 GPa (filled symbols). 
Squares correspond to $^3$He and circles to $^4$He.
For each pressure, results are displayed for temperatures at which the 
considered solids were found to be stable along the PIMC simulations.
As expected, $a$ is larger for $^3$He than for $^4$He, and the difference
$\delta a = a_3 - a_4$ is smaller for higher pressure. 
Also, for a given pressure, $\delta a$ decreases slowly as the temperature is
raised (at higher $T$ the solid becomes ``more classical").

\begin{figure}
\vspace{-2.0cm}
\includegraphics[width= 9cm]{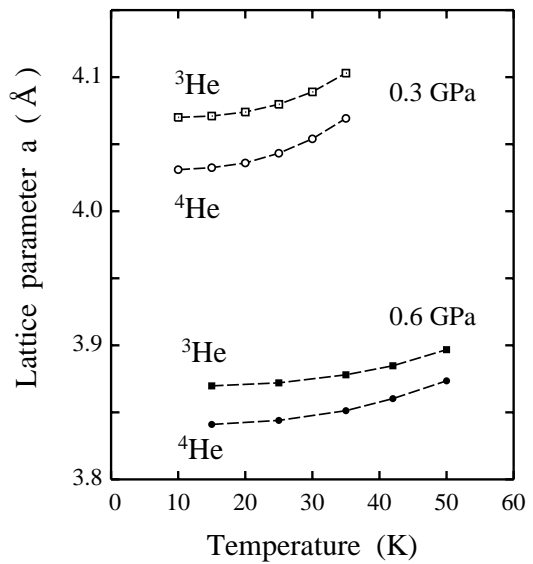}
\vspace{-2.5cm}
\caption{\label{f2}
Temperature dependence of the lattice parameter $a$ of fcc helium, derived
from PIMC simulations at two different pressures: 0.3 and 0.6 GPa.
Squares and circles indicate results for $^3$He and $^4$He, respectively.
Error bars are smaller than the symbol size.
Dashed lines are guides to the eye.
}
\end{figure}

To quantify the change in crystal volume with isotopic mass, we calculate
the ratio $\delta V / V_4 = (V_3 - V_4) / V_4$. This ratio is shown in
Fig.~3 as a function of pressure for fcc (filled symbols) and hcp He 
(open symbols) at 25 K.
At $P$ = 0.2 GPa, we find $\delta V / V_4$ = 0.030. This value is reduced
by more than a factor of 3 at $P$ = 12 GPa, where we obtain 
$\delta V / V_4 = 7.7 \times 10^{-3}$.
It is interesting to note that at $P$ = 1 GPa, both fcc and hcp phases could
be simulated at 25 K, remaining (meta)stable along the corresponding simulation
runs.
The ratio $\delta V / V_4$ obtained in both cases is shown in Fig.~3, and
the results coincide within statistical errors.
Direct experimental measurements of this isotopic effect on the molar volume
are scarce. Stewart \cite{st63} measured the molar volumes of both $^3$He
and $^4$He at 4.2 K and pressures up to 2 GPa. In particular, for $P$ = 0.3
and 0.6 GPa he found $V_4 / V_3$ = 1.023. At these pressures, we found for
the low-temperature limit in the PIMC simulations $V_4 / V_3$ = 1.026(1)
and 1.023(1), respectively.  

\begin{figure}
\vspace{-2.0cm}
\includegraphics[width= 9cm]{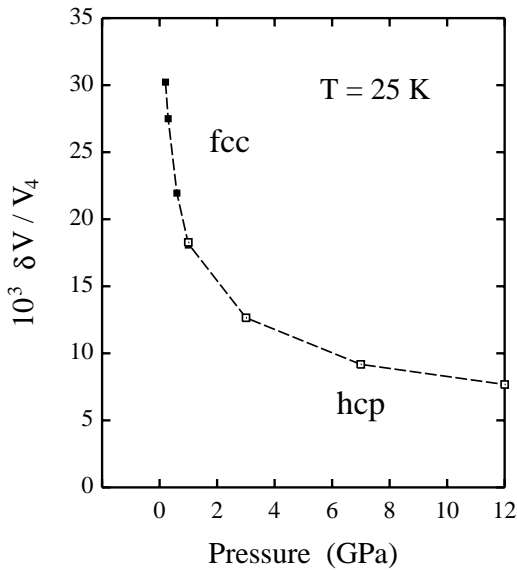}
\vspace{-2.5cm}
\caption{\label{f3}
Isotopic effect on the crystal volume of solid helium, as obtained from
PIMC simulations. Shown is the ratio $\delta V / V_4 = (V_3 - V_4) / V_4$ as
a function of pressure at 25 K.
Open and filled symbols correspond to hcp and fcc phases, respectively.
Error bars are less than the symbol size. The dashed line is a guide
to the eye.
}
\end{figure}

 The difference $\delta V$ is largest at small pressures and low
temperatures, where quantum effects are most important. 
For a given solid, quantum effects on the crystal volume can be measured
by the difference $V - V_{\rm cl}$
between the actual volume $V$ and that obtained for
a ``classical'' crystal of point particles, $V_{\rm cl}$.
This difference decreases for increasing atomic mass and 
temperature \cite{mu95,he01}.
From our PIMC simulations at $T$ = 25 K and a relatively low pressure
of 0.3 GPa, we found an increase in the volume of solid $^3$He and
$^4$He of 26\% and 22\% respectively, as compared to the ``classical'' 
crystal at zero temperature.

PIMC simulations allow us to obtain the kinetic energy of the different
atoms in the simulation cell at finite temperatures.
For pure fcc crystals of $^3$He and $^4$He, we find a kinetic energy $E_k=$
9.22 and 8.21$\pm 0.03$ meV/atom, respectively (at 25 K and 0.3 GPa).
This translates into a kinetic-energy ratio of 1.12, slightly lower than the 
ratio between zero-point energies expected in a harmonic approximation
($E_0^3 / E_0^4 = \sqrt{4/3} = 1.15$). 
These energy values are similar to those obtained earlier from PIMC simulations
in the $NVT$ ensemble. In particular, our results for $^4$He 
(giving a molar volume of 9.95 cm$^3$/mol) are in line with those obtained
earlier in the $NVT$ ensemble at temperatures close to 25 K
and molar volumes around 10 cm$^3$/mol \cite{dr00}, which in turn
agree with experimental measurements \cite{di04}.

For isotopically-mixed crystals, one expects the kinetic energy $E_k$ of the
whole crystal to evolve smoothly as a function of the mean isotopic
mass $\overline{M}$.   
In Fig.~4 we show the kinetic energy of fcc helium vs $\overline{M}$ at 
25 K and 0.3 GPa, as derived from our PIMC simulations. 
Within the precision of our results, we observe a linear dependence
of $E_k$ vs $\overline{M}$, as indicated by the dashed line.
We have compared these results with those obtained from PIMC simulations 
for the virtual crystal with mass $\overline{M}$, and found that 
differences between both sets of results are smaller than the statistical 
noise. A discussion on the relation between vibrational kinetic and potential
energies in solid helium was given elsewhere \cite{he06}, and will not 
be repeated here.

\begin{figure}
\vspace{-2.0cm}
\includegraphics[width= 9cm]{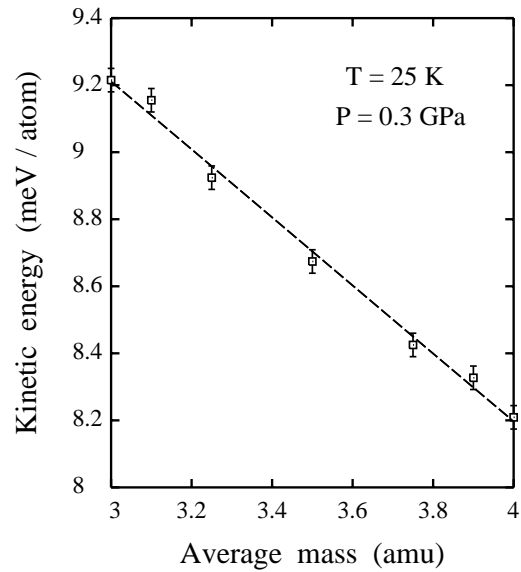}
\vspace{-2.5cm}
\caption{\label{f4}
Kinetic energy of fcc helium as a function of the mean isotopic mass
$\overline{M}$ in isotopically mixed crystals. Symbols are results of
PIMC simulations at 25 K and 0.3 GPa.
The dashed line is a linear fit to the data points.
}
\end{figure}

Going back to the lattice parameter of fcc He, in Fig.~5 we display the 
dependence of $a$ on the average isotopic mass, at 25 K and 0.3 GPa.
Open squares correspond to simulations in which $^3$He and $^4$He
atoms were randomly distributed over the crystal sites, according to
the required average mass $\overline{M}$.
The results show a linear dependence of $a$ on $\overline{M}$.
In Fig.~5 we also show results of PIMC simulations of fcc He, where
each atom has a fictitious mass $\overline{M}$
($M_i = \overline{M}$ for $i = 1, ..., N$).
This virtual-crystal approximation yields for $\overline{M}$ different
from 3 and 4 amu a lattice parameter smaller than that obtained for a 
realistic distribution of isotopes over the simulation cell.
For an isotopic mixture including 50\% of each isotope, we find
$a$ = 4.06162 \AA\ vs $a_{\rm vc}$ = 4.05950 \AA\ for the virtual-crystal
approximation, i.e., a difference between both models 
$a - a_{\rm vc} = 2.12 \times 10^{-3}$ \AA. For this composition we have 
considered 15 different isotope distributions, and found for the lattice 
parameter a standard deviation $\sigma = 2.5 \times 10^{-4}$ \AA, 
a little smaller than the symbol size in Fig.~5. Thus, the difference
$a - a_{\rm vc}$ amounts to about $8 \sigma$.
For 75\% $^3$He, we took 12 realizations of the isotope mixture,
and found $a - a_{\rm vc} = 1.64 \times 10^{-3}$ \AA. In this case, 
$\sigma = 2.8 \times 10^{-4}$ \AA, or $a - a_{\rm vc} \approx 6 \sigma$.

\begin{figure}
\vspace{-2.0cm}
\includegraphics[width= 9cm]{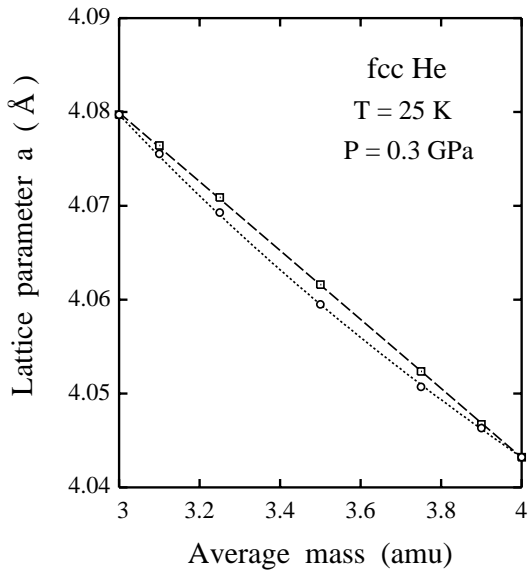}
\vspace{-2.5cm}
\caption{\label{f5}
Lattice parameter $a$ of fcc helium as a function of the mean isotopic mass
$\overline{M}$. Data points were obtained from PIMC simulations at 25 K and
0.3 GPa, for isotopically mixed crystals containing $^3$He and $^4$He
(squares), and for virtual crystals built up by atoms with the average
isotopic mass (circles).
Error bars are smaller than the symbol size.
The dashed line is a least-square fit to the data points (squares).
The dotted line was obtained from a quasi-harmonic approximation using
Eq. (5).
}
\end{figure}

\section{Discussion}

Path-integral Monte Carlo simulations have been found to be well suited to
study finite-temperature anharmonic effects on structural and thermodynamic
properties of crystalline solids. These effects are particularly important
for solid helium, where
isotopic effects are relevant, as manifested in differences in the molar 
volume and vibrational energies of solid $^3$He and $^4$He.
The PIMC method enables us to study phonon-related properties without
the assumptions of quasiharmonic or self-consistent phonon approximations,
and to study anharmonic effects in solids in a non-perturbative way.
Thus, for a given reliable interatomic potential, this method
yields in principle ``exact'' values for measurable
properties of many-body quantum problems, with an accuracy limited by the
imaginary-time step (Trotter number) and the statistical uncertainty of the 
Monte Carlo sampling.

Our results for the lattice parameter of solid helium as a function
of the average isotopic mass (see Fig.~5) can be understood in terms of the 
theory of alloys. In fact, $^3$He and $^4$He behave in this respect as atoms 
with different atomic radii, as a consequence of the different vibrational
amplitudes of both isotopes. If we consider a $^4$He crystal with
$^3$He impurities, the lattice expansion $\Delta a$ due to these impurities is
given by Vegard's law \cite{ve21,de91}:
\begin{equation}
       \frac{\Delta a}{a_4} = \beta C_3   \; ,
\label{vegard}
\end{equation}
where $C_3$ is the concentration of $^3$He.
From the data presented in Fig.~5 we find an expansion coefficient 
$\beta = 1.49 \times 10^{-25}$ cm$^3$/atom.
According to our results, this linear relation between lattice parameter 
and isotope concentration holds for the whole concentration range,
as shown in Fig.~5 (squares and dashed line). 
This is in agreement with earlier calculations for alloys, which indicate that 
for a difference between atomic radii less than $\sim 5\%$ one does
not expect appreciable departure from linearity \cite{de91}. In fact, 
$^3$He and $^4$He behave as atoms with a difference in 
effective atomic radii of about 1\% (for $T$ and $P$ considered here).

The virtual-crystal approximation has been employed earlier to study
the dependence of lattice parameters on the average isotopic mass. 
This can be conveniently done by using a
quasiharmonic approach, according to which the low-temperature lattice 
parameter $a$ for mean isotopic mass $\overline{M}$ can be approximated 
by \cite{de96}
\begin{equation}
a = a_{\infty} + \frac{1}{6 B a^2_{\infty}}
   \sum_{n, {\bf q}}  \hbar \omega_n({\bf q}) \gamma_n({\bf q})
    \hspace{0.2cm}   .
\label{am}
\end{equation}
Here, $\omega_n({\bf q})$ are the frequencies of the $n$th
mode in the crystal, $B$ is the bulk modulus,
$a_{\infty}$ is the zero-temperature lattice constant
 in the limit of infinite atomic mass (classical limit), and
$\gamma_n({\bf q}) = - \partial \ln \omega_n({\bf q}) /
 \partial \ln V $ is the Gr\"uneisen parameter of mode
$n, {\bf q}$.
Assuming a mass dependence of the frequencies 
$\omega_n({\bf q}) \sim \overline{M}^{-1/2}$, one finds for the relative 
change in lattice parameter with isotopic mass
\begin{equation}
a  \approx  a_{\infty} + A  \overline{M}^{-\frac12} \; ,
\label{am2}
\end{equation}
where $A$ is constant for a given pressure.
By applying this equation to $a_3$ and $a_4$, we can find $A$, 
and then the dependence of $a$ on $\overline{M}$ shown 
in Fig.~5 as a dotted line.
This line coincides within error bars with the results derived from
the PIMC simulations for helium with $M_i = \overline{M}$ for all $i$ 
(virtual crystal).
Note that the low-temperature expression for the lattice parameter given 
in Eq. (4) is a good approximation for the conditions ($P, T$) considered
here. In fact, for a volume of $\sim$ 10 cm$^3$/mole, a temperature of 25 K
can be considered a ``low temperature'' when compared with 
$\Theta_D \gtrsim$ 120 K \cite{ar03}.    

It is worth commenting on the difference found between this quasiharmonic 
approximation and the actual isotope distribution over the crystal. 
For the virtual crystal we found in both PIMC simulations and in the
quasiharmonic approach, a non-linear dependence of the lattice parameter
$a$ on average isotopic mass. This contrasts with the linear dependence
derived from PIMC simulations for actual distributions of isotopes.
(If a departure from linearity appears in this case, it will be smaller
than the precision of our results.) This observation is important for
the helium solid solutions considered here, since in most known solids
both approaches give the same results \cite{he01b,he03a,de96}.
From the results displayed in Fig.~5, we observe for $\overline{M}$ = 3.5 amu
a difference $a - a_{\rm vc} = 2.1 \times 10^{-3}$ \AA, which amounts to about 
0.05\% of the lattice parameter. This relative difference is much larger
than the uncertainty in structural parameters currently derived from 
diffraction methods \cite{ka98}.

In the last few years, several authors have indicated that pressure causes 
a decrease in anharmonicity \cite{ka03,la04,he05}, in agreement with earlier 
observations that
the accuracy of quasiharmonic approximations increases as pressure is raised
and the density of the solid increases \cite{po72,ho73}. 
This is also the origin of the decrease in the isotopic effects studied
here, as pressure is raised. Since these isotopic effects are caused by
anharmonicity of the interatomic interactions, an effective decrease in
anharmonicity causes a reduction in the difference $\delta a = a_3 - a_4$
or, equivalently, in the ratio $\delta V / V_4$ shown in Fig.~3. 

In summary, we have carried out PIMC simulations of solid solutions of helium 
isotopes in the isothermal-isobaric ensemble. Our results indicate that 
Vegard's law is fulfilled in the whole composition range, i.e., the crystal
volume changes linearly with isotopic composition. This volume change
decreases appreciably as pressure is raised, but it is still clearly 
observable at pressures of the order of 10 GPa. Approximations such as
virtual crystals with average atomic mass are not valid for solid isotopic 
helium mixtures.

\begin{acknowledgments}
The author benefitted from discussions with R. Ram\'{\i}rez.
This work was supported by Ministerio de Educaci\'on y Ciencia  (Spain) 
through Grant No. FIS2006-12117-C04-03.   \\
\end{acknowledgments}

\end{document}